\documentclass[superscriptaddress,showkeys,prl,aps,twocolumn]{revtex4-1}
\usepackage{dcolumn}% Align table columns on decimal point
\usepackage{bm}% bold math

\usepackage[colorlinks,hyperindex, urlcolor=blue, linkcolor=blue,citecolor=black, linkbordercolor={.7 .8 .8}]{hyperref}
\usepackage{graphicx}
\usepackage{float}
\usepackage{amsfonts}
\usepackage{amsmath}
\usepackage{amssymb}
\usepackage{amsbsy}
\usepackage{multirow}
\usepackage{nicefrac}
\usepackage{siunitx}
\usepackage{caption}
\begin{document}

\title {Hall field-induced magneto-oscillations near charge neutrality point in graphene}

\author{Mrityunjay Pandey}
\email{mrityunjay@iisc.ac.in}
\address{Centre for Nano Science and Engineering, Indian Institute of Science, Bangalore 560012, India}

\author{Kenji Watanabe}
%\email{watanabekenji@nims.go.jp}
\address{Research Center for Functional Materials, National Institute for Materials Science, 1-1 Namiki, Tsukuba 305-0044, Japan}
\author{Takashi Taniguchi}
%\email{taniguchitakashi@nims.go.jp}
\address{International Center for Materials Nanoarchitectonics, National Institute for Materials Science, 1-1 Namiki, Tsukuba 305-0044, Japan}
\author{Srinivasan Raghavan}
%\email{sraghavan@iisc.ac.in}
\address{Centre for Nano Science and Engineering, Indian Institute of Science, Bangalore 560012, India}
\author{U. Chandni}
\email{chandniu@iisc.ac.in}
\address{Department of Instrumentation and Applied Physics, Indian Institute of Science, Bangalore 560012, India}

\begin{abstract}
 
We explore the non-equilibrium transport regime in graphene using a large dc current in combination with a perpendicular magnetic field. The strong in-plane Hall field that is generated in the bulk of the graphene channel results in Landau levels that are tilted spatially. The energy of cyclotron orbits in the bulk varies as a function of the spatial position of the guiding center, enabling us to observe a series of compelling features. While  Shubnikov-de Haas oscillations are predictably suppressed in the presence of the Hall field,  a set of fresh magnetoresistance oscillations emerge near the charge neutrality point as a function of dc current. Two branches of oscillations with linear dispersions are evident as we vary carrier density and dc current, the velocity of which closely resembles the TA and LA phonon modes, suggestive of phonon-assisted intra-Landau level transitions between adjacent cyclotron orbits. Our results offer unique possibilities to explore non-equilibrium phenomena in two-dimensional materials and van der Waals heterostructures.\\

\end{abstract}

%\date{\today}

\maketitle
%\section{I. Introduction}
In two dimensional electron systems, magneto oscillations underpin the Landau quantization of electrons under an external magnetic field. In addition to the conventional Shubnikov-de Haas (SdH) oscillations, a variety of novel low-field oscillations have been investigated, primarily in GaAs-AlGaAs heterostructures, where transition of electrons between Landau levels can lead to oscillations in longitudinal resistance as the magnetic field ($B$) is varied. These transitions are typically enabled by phonon-mediated resonances~\cite{C3_003,C3_001,C3_002,C3_006,C3_14,C3_0018}, strong Hall field~\cite{C3_0018,C3_17,C3_18,C3_005} or microwave illumination~\cite{C3_39,C3_40,C3_41,C3_42,C3_0017}. In recent times, quantum oscillations have led to key discoveries in graphene and other van der Waals heterostructures, including unconventional Berry's phases~\cite{C3_01,C3_02}, the Hofstadter spectrum~\cite{C3_03} and  magnetophonon resonances~\cite{C3_34}. These studies were primarily done in the near-equilibrium limit with a small electric field applied laterally. The fate of quantum Hall effect and magneto oscillations in graphene when subjected to strong electric fields and high currents has received much less attention~\cite{C3_008,C3_009,C3_0012,C3_0013}. Importantly, such non-equilibrium effects are relevant in the context of electron interferometers using edge channels for quantum information processing~\cite{C3_19,C3_21}, generation of excitations such as magnetoplasmons in the edge channels~\cite{C3_29} and for the application of the quantum Hall effect as a resistance standard~\cite{ C3_0013,C3_0024}.\\
 \begin{figure*}

\includegraphics[width=5in]{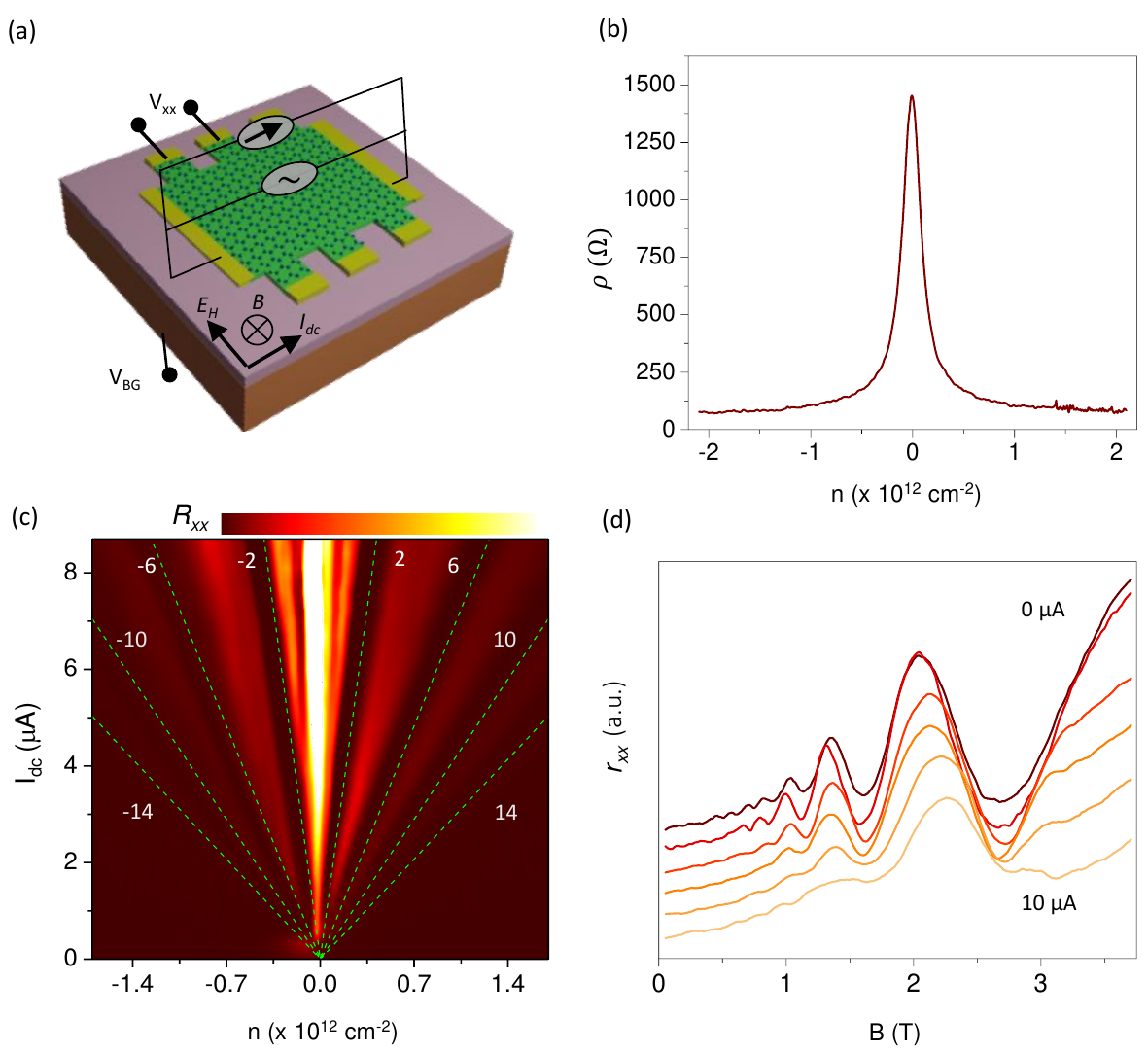}
%\captionsetup{labelformat=empty}
\captionsetup{justification=raggedright,singlelinecheck=false}
%\justify
\caption{\textbf{Electrical characterization:} (a) Schematic of the non-equilibrium magneto-transport measurement on a Si/SiO$_{2}$ back gated device with a width of 18 $\mu$m and voltage probes separated by 3.8 $\mu$m where V$_{BG}$ is the back gate voltage and $V_{xx}$ is the longitudinal voltage drop. The direction of magnetic field ($B$), dc current ($I_{\textit{dc}}$), and Hall field ($E_{H}$) are mutually perpendicular to each other. (b) Longitudinal resistivity $\rho$ as function of charge carrier density $n$ measured at 2.5 K for the graphene Hall bar device.  (c) Landau fan diagram of the longitudinal resistance $R_{xx}$ as a function of $n$ and $B$, with the dashed lines denoting filling factors.  Scale bar extends from 0.25 \si{\ohm} (red) 40 k\si{\ohm} (yellow). (d) Differential resistance \textit{r$_{xx}$} as a function of $B$ for  I$_{\textit{dc}}$ ranging between 0 to 10 $\mu$A in steps of 2 $\mu$A shows suppression of SdH oscillations on increasing $I_{dc}$. Data taken at a magnetic field between 0 to 3.75 T at $n$ = 3.5 x 10$^{11}$ cm$^{-2}$}
\end{figure*}

\begin{figure*} 
               \includegraphics[width=7in]{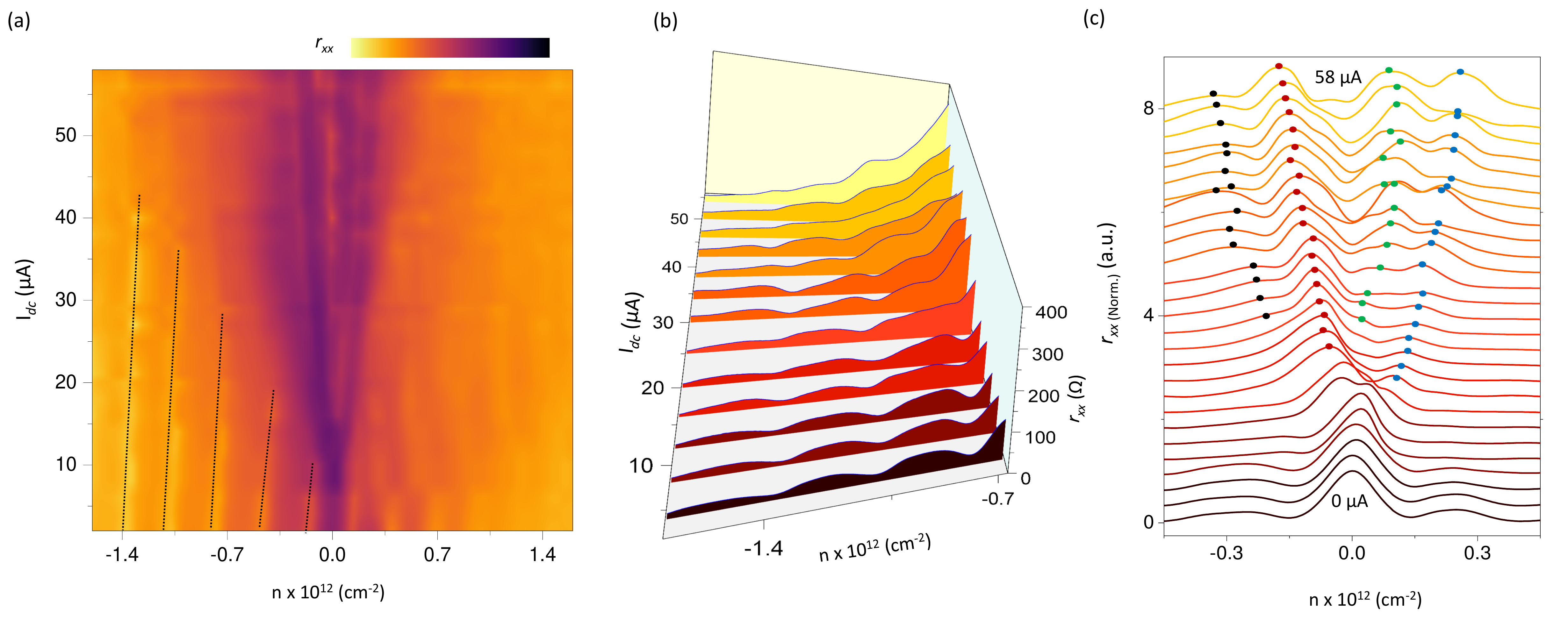}
               %\captionsetup{labelformat=empty}
               \captionsetup{justification=raggedright,singlelinecheck=false}
             % \justify 
              \caption{\textbf{Non-equilibrium measurements at constant \textit{B}:} (a) Differential resistance \textit{$r_{xx}$} at a fixed magnetic field of 3 T as function of $n$ and $I_{dc}$, at 2.5~K,  with $I_{\textit{dc}}$ ranging from 0 $\mu$A to 58 $\mu$A. Color map is plotted in log scale. Black dashed line indicate the resistance minima corresponding to LL indices $\nu$ = 18, 14, 10, 6 and 2. Scale bar ranges from 0.25 \si{\ohm} (yellow) to 6.4 k\si{\ohm} (purple). (b) \textit{$r_{xx}$} at different $I_{\textit{dc}}$ for the hole side showing the rise in the background resistance, as well as the resistance minima corresponding to various LL indices.  (c) Normalised resistance \textit{$r_{norm}$} shown as line plots near the charge neutrality point elucidating the appearance of a set of new oscillations as a function of $I_{dc}$. Individual line plots are offset in Y-axis for clarity. Peak positions are marked using dots.}
               \end{figure*}

In this letter, magnetotransport in the non-equilibrium regime is investigated in a large area graphene Hall bar device, by introducing a high dc current ($I_{\textit{dc}}$) in addition to a small ac excitation (see Fig.1(a)).  The high quality of our hexagonal boron nitride (hBN)-encapsulated device allows us to probe a series of phenomena over a wide range of charge carrier densities. We observe signatures of archetypal charge carrier heating-induced dissipation in differential resistance measurements, that are consistent with the presence of a Hall field ($E_H$). In the presence of $I_{dc}$, to a good approximation, the drift velocity $v_d$ due to $E_H$ can be expressed as   $v_{d} = I_{dc}/new$, where \textit{w} is the width of the channel, \textit{n} is the carrier density and \textit{e} is the electronic charge. In the regime \textit{v$_{d}$} $<$ \textit{v$_{p}$}, $\omega_c/2k_F$ (where  $v_p$ is the phonon velocity, $\omega_c$ is the cyclotron frequency and $k_F$ is the Fermi wave vector), we observe a marked suppression of SdH oscillations at constant $n$ when $I_{dc}$ is increased, signifying carrier heating. On the other hand, quantum oscillations show several interesting features when $n$ is varied at fixed $B$. The channel shows dissipative transport as $I_{dc}$ is increased, with the quantum oscillations disappearing in the low-$n$ regions at a lower $I_{dc}$ in comparison to the high-$n$ regions. Most strikingly, near the charge neutrality point we report a set of novel resistance oscillations, that disperse linearly as a function of $I_{dc}$ and $n$. These oscillations are observed in the density regime where $v_d\approx v_p$. The two sets of oscillations observed in our data resemble the LA and TA modes of acoustic phonons in graphene, providing evidence for Hall field-induced inelastic intra-Landau level transitions near the charge neutrality point. \\

The experiments reported in this study were performed at 2.5 K on a hBN-encapsulated graphene Hall bar device with a width of 18 $\mu$m. It was shown in recent studies~\cite{C3_34,C3_351} that devices with larger widths in comparison to the mean free path of carriers can provide a platform to investigate the resistive behaviour of the bulk by minimizing edge contributions. This regime is relevant to our devices as well, where the electron-phonon interactions are found to be crucial. Fig. 1(b)-(c) show the basic electrical characterization of our device at $I_{dc}=0$.  Fig. 1(b) shows the four-probe resistivity as a function of the charge carrier density $n$. Fig. 1(c) indicates the Landau fan diagram in the $(n,B)$ parameter space, with filling factors $\nu$ = $\pm$ 2, 6, 10 and 14 indicated by dashed lines. SdH oscillations are investigated at several $I_{dc}$, as shown in Fig. 1(d), where the differential resistance \textit{r$_{xx}$} as a function of $B$ for several  $I_{dc}$ ranging from 0 to $10~\mu$A are shown for a carrier density of 0.35 x 10$^{12}$ cm$^{-2}$. The high quality of our sample is apparent from the observation of SdH oscillations at a low field of 0.5 T. The oscillations were found to be suppressed with increase in $I_{dc}$. We have studied the carrier energy loss to the lattice by measuring the SdH oscillations and analyzing the damping of these oscillations (See supplementary information). In graphene, the rate of energy loss is given as:
 \begin{equation}
P = I_{dc}^{2}r_{xx}/n A
 \end{equation}
 where $P$ is the dissipation power and \textit{A} is the area of the channel, indicating the power loss is expectedly large at low density and large current~\cite{C3_30,C3_31}. Consequently, carrier temperature rises with increasing current and decreasing carrier density, leading to dissipative transport in the channel, consistent with our experimental observations. The estimated dissipation rates indicate that the energy loss rate at $I_{dc}=10~\mu$A is of order 10$^{-14}$ W however, at 2 $\mu$A it is of order 10$^{-16}$ W, resulting in high carrier temperatures at large dc currents as reported in previous studies~\cite{C3_30,C3_31}.  This estimate assumes a very rapid thermalization of the electrons on a time scale of few tens of femtoseconds, corroborated by recent experiments in graphene and related materials \cite{C3_32, C3_33}
   
 \begin{figure}
         \includegraphics[scale=0.5]{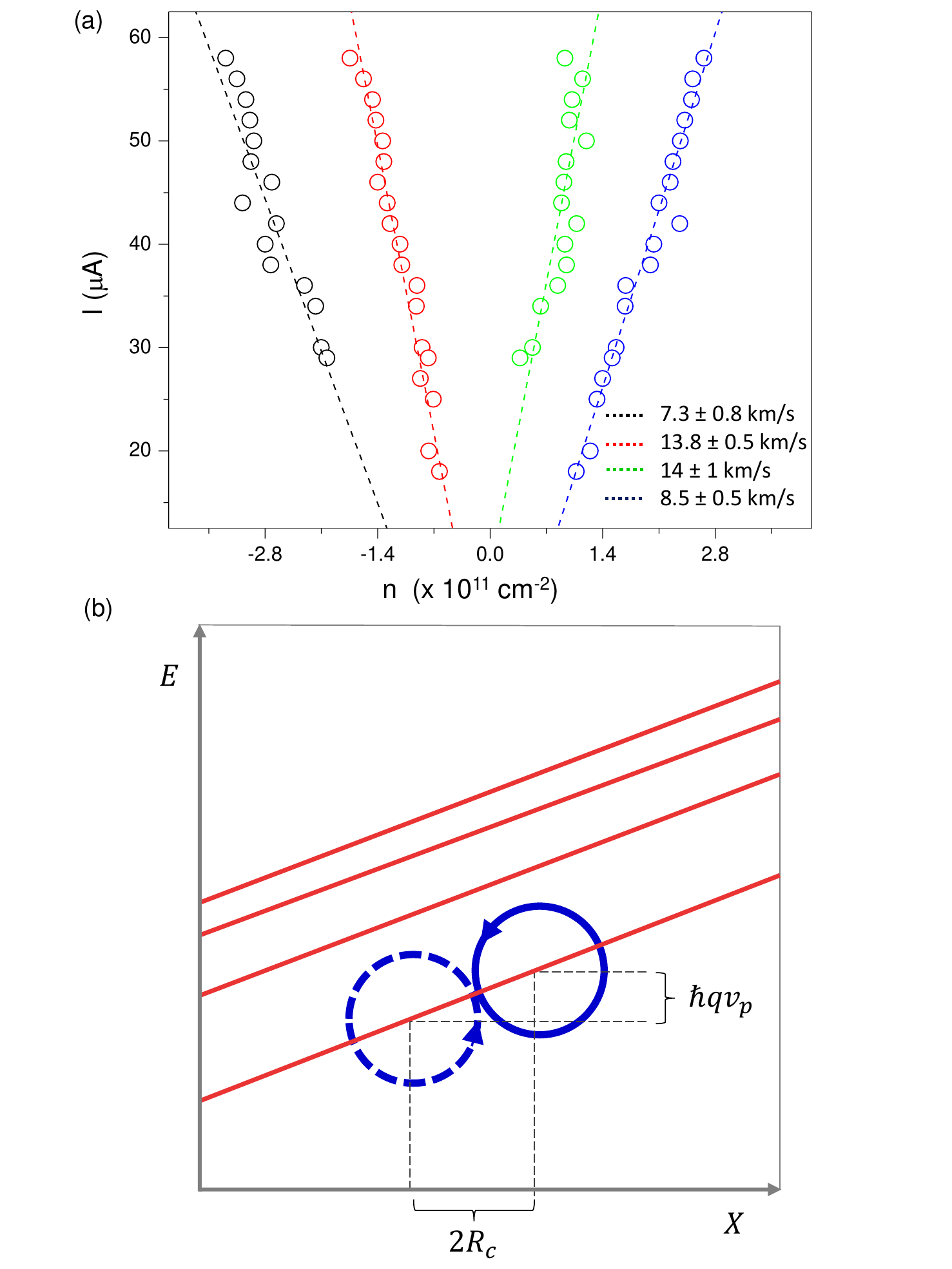}
         %\captionsetup{labelformat=empty}
         \captionsetup{justification=raggedright,singlelinecheck=false}
        % \justify
        \caption{\textbf{Intra-Landau level transitions near the charge neutrality point:} (a) Linear fitting of the resistance peaks denoted by dots in Fig.2(c) near the charge neutrality point, as function of $n$ and $I_{dc}$.  Velocities calculated from the slopes are 14 and 8.5 km/s on the electron side and 13.8 and 7.3 km/s on the hole side, respectively, which match closely with the LA and TA phonon modes in graphene. (b) Schematic showing the possible mechanism of intra Landau level transitions in presence of a Hall field, assisted by acoustic phonons. In presence of a Hall field, Landau levels are tilted spatially as denoted by the red lines. A figure-of-8 configuration occurs for a separation of $2R_c$, leading to maximum overlap of wavefunctions and transitions between adjacent cyclotron orbits, favoured by phonons of energy $\hbar qv_p$.}
\end{figure}

Our most significant results are illustrated in Fig. 2.  In Fig. 2a, $r_{xx}$ is plotted as a color map for several $n$ and $I_{dc}$ values, at a fixed magnetic field of 3 T. At $I_{dc}=0$, resistance minima corresponding to filling factors $|\nu|= 2$, 6, 10, 14 and 18 are clearly visible. As $I_{dc}$ is increased, the resistance value at the minima increases and eventually merge with the higher resistance background. This feature is clearer on the hole side denoted by the dashed lines. Remarkably, the resistance minima for higher $\nu$ sustain till a larger $I_{dc}$. This observation is consistent with $I_{dc}$-induced carrier heating. Electron temperature increases with the increase in $I_{dc}$ resulting in disappearance of the resistance minima. Energy loss rate as a function of $n$ at different $I_{dc}$ calculated using eq. 1, reveals that dissipation rate is larger at low density. Consequently, the quantum oscillations are suppressed at a lower $I_{dc}$ for lower $n$. This also results in an increase in carrier temperature and the overall resistance background as depicted in Fig. 2(b), which shows a sectional cut of the color map in Fig. 2(a).

The most striking observation occurs near the charge neutrality point, where we observe a new set of resistance oscillations to emerge, which become increasingly stronger as $I_{dc}$ is increased. Fig. 2(c) shows the section of magneto resistance (normalized as $r_{norm}=r_{xx}/r_{xx}^o$, where $r_{xx}^o$ is the differential resistance at the charge neutrality point) in the range of $\pm$ 0.4 x 10$^{12}$ cm$^{-2}$.  It is evident from Fig. 2(c) that there are four branches of resistance oscillations, two each on the electron and hole sides, which show dispersive behavior over the ($n-I_{dc}$) subspace. Notably, these oscillations were observed to be broader at a higher magnetic field of 5 T (see supplementary information), while retaining the general characteristics. 

Magneto oscillations in two dimensional electron systems can have varied origins depending on the experimental configurations. For instance, inter-Landau level transitions are known to be favoured in microwave-induced resistance oscillations~\cite{C3_42,C3_0017}, Zener tunneling at high currents~\cite{C3_0018,C3_17,C3_18} and magnetophonon oscillations involving acoustic or optical phonons~\cite{C3_001,C3_14}. In all of these, the transitions occur via a large momentum transfer of 2$k_F$, where $k_F$ is the Fermi wave number, which is equivalent to a guiding center shift of twice the cyclotron radius or 2$R_c$. Semiclassically, this implies that the transitions between Landau levels are strongest when there is maximum spatial overlap of wavefunctions in the figure-of-8 configuration for the orbits. This momentum is provided by short range scatterers, phonons or defects in the sample. Particularly in case of experiments involving non-equilibrium transport, the Landau levels are tilted by the Hall field, resulting in resistance oscillations when the energy and momentum conservation rules are satisfied \textit{viz.} $j\hbar \omega_c=2\hbar k_F v_F$, where $j$ is an integer. However, such transitions across different Landau level orbits would need a large Hall field to be present across a length scale of $2R_c$, particularly at the lowest Landau levels in graphene which have the highest energy seperation. In our experiment, the oscillations near the vicinity of the charge neutrality point appear around the zeroth Landau level, where $2R_ceE_H<j\hbar \omega_c$, restricting inter Landau level transitions. Hence, in the Hall field regime that we have explored, the magneto oscillations near charge neutrality point are rather surprising and cannot be understood based on the above mechanisms. 

We restrict our analysis to intra-LL transitions in light of the discussion above. Since the Hall field tilts the Landau levels, it is imperative to have an external excitation to aid the transitions we observe, since the energies of the orbits at finite $E_H$ are  not uniformly equal to the Landau level energy at $E_H=0$. It is unlikely that the optical phonons contribute to this process since we observe a dispersive feature in the ($n-I_{dc}$) space. Magneto-excitons are also improbable candidates~\cite{C3_0021,C3_00210,C3_00211} since they involve electron-hole pairs in adjacent cyclotron orbits belonging to different Landau levels.  To further understand the nature of dispersion and the excitations underlying the resistance oscillations, we have estimated the Hall field-induced drift velocity as a function of $n$ and $I_{dc}$ (see supplementary figure S2). 
. 
It is apparent that the novel oscillations observed in the magneto resistance at low densities in Fig. 2 occur when the drift velocity becomes nearly equal to the velocities of acoustic phonons in graphene ($\sim 10$ km/s), enabling phonon-assisted transitions between adjacent cyclotron orbits belonging to the same Landau level.

 To extract the velocity of the relevant phonon excitations, the peak positions as a function of $I_{\textit{dc}}$ and \textit{n} are plotted in Fig. 3(a) and the velocity is estimated as $v_{d} = I_{dc}/new$. The drift velocities associated with the two branches were found to be $7.3\pm 0.8$ km/s and $13.8\pm 0.5$ km/s on the hole side and $8.5 \pm 0.5$ km/s and $14 \pm 1$ km/s on the electron side, respectively,  which match closely with the LA and TA phonon modes in graphene~\cite{C3_43,C3_0025}. We speculate that charge inhomogenieties near the neutrality point may lead to a marginal decrease in the measured phonon velocities. Fig. 3(b) summarizes the mechanism discussed above. In the presence of a Hall field \textit {$E_{H}$}, cyclotron orbits of an index \textit{N} are no longer of equal energy due to the spatial tilting of Landau levels given by the relation
   \begin{equation}
   E_{N}(k_{y}) = E_{N}- eE_{H}X(k_{y})
   \end{equation} 
  where $E_N$ denotes the energy of the $N$th Landau level and $X(k_y)=-k_yl_B^2$; $k_y$ is the wavevector along $Y-$axis and $l_B$ is the magnetic length. Consequently, the guiding centers of the adjacent cyclotron orbits are at different energies.  Transitions are possible within a Landau level when the energy shift in the guiding centers of adjacent orbits $eE_H2R_c$ matches the phonon energy  \textit{$\hbar$qv$_{p}$}, with $q=2k_F$ as shown in Fig. 3(b), which is equivalent to the condition $v_d=v_p$.

\subsection{ Conclusions}
In summary,  our study of non-equilibrium transport in large area graphene devices provide the framework to explore the strong Hall field regime. We report pronounced Hall-field induced resistance oscillations near the charge neutrality point in graphene, in addition to dissipative carrier heating effects on Landau levels. It is apparent that these novel oscillations have a linear dispersion with $I_{dc}$ and $n$, indicative of electron-phonon interactions and intra-Landau level transitions. Remarkably, Hall-field induced oscillations also provide a novel method to investigate low-energy phonon modes in graphene and other two-dimensional layered materials and heterostructures. 

\subsection{ Methods}
Large area, high-quality heterostructures of hBN/Graphene/hBN were fabricated using a dry pick up and transfer technique using a polymer stack of PPC/PDMS. The stack was transferred on a Si/SiO$_2$ wafer and annealed for 4 hours in argon at 2 SLM flow rate at 350$^{\circ}$C. The geometrical extent of the large encapsulated graphene was identified using electrostatic force microscopy ~\cite{C3_36} to choose a clean region of interest. Hall bar was etched using ebeam lithography and reactive ion etching, and Cr/Au (5/50 nm) contacts were defined using standard device fabrication methods. All the electrical measurements were performed in a Janis cryostat at 2.5 K. SR-830 lock-in amplifier was used at 13 Hz to provide a small ac excitation current of 40 nA, while a dc current was introduced using a Kepco programmer. An isolation transformer circuit was used to pass both ac and dc currents through the Hall bar channel. Keithley 2450 source meter was used to source the back gate voltage.

\subsection{Acknowledgements}
 We thank S. Banerjee, J. P. Eisenstein, J. Jung, H. R. Krishnamurthy, J. Song and J. Sutradhar for useful discussions and inputs. We gratefully acknowledge the usage of the Micro and Nano Characterization Facility (MNCF) and National Nanofabrication Centre at CeNSE, IISc. U.C. acknowledges support from Science and Engineering Research Board (SERB) via grant  SPG/2020/000164. KW and TT acknowledge support from the Elemental Strategy Initiative conducted by the MEXT, Japan (Grant Number JPMXP0112101001) and JSPS KAKENHI (Grant Numbers JP19H05790 and JP20H00354).
 
\bibliographystyle{achemso}
\bibliography{reference}

\newpage
\clearpage
\onecolumngrid
\begin{center}
\textbf{\Large{Supporting Information}}
\end{center}
\maketitle

    \begin{figure*}[h]
             \centering
              \includegraphics[scale=0.7]{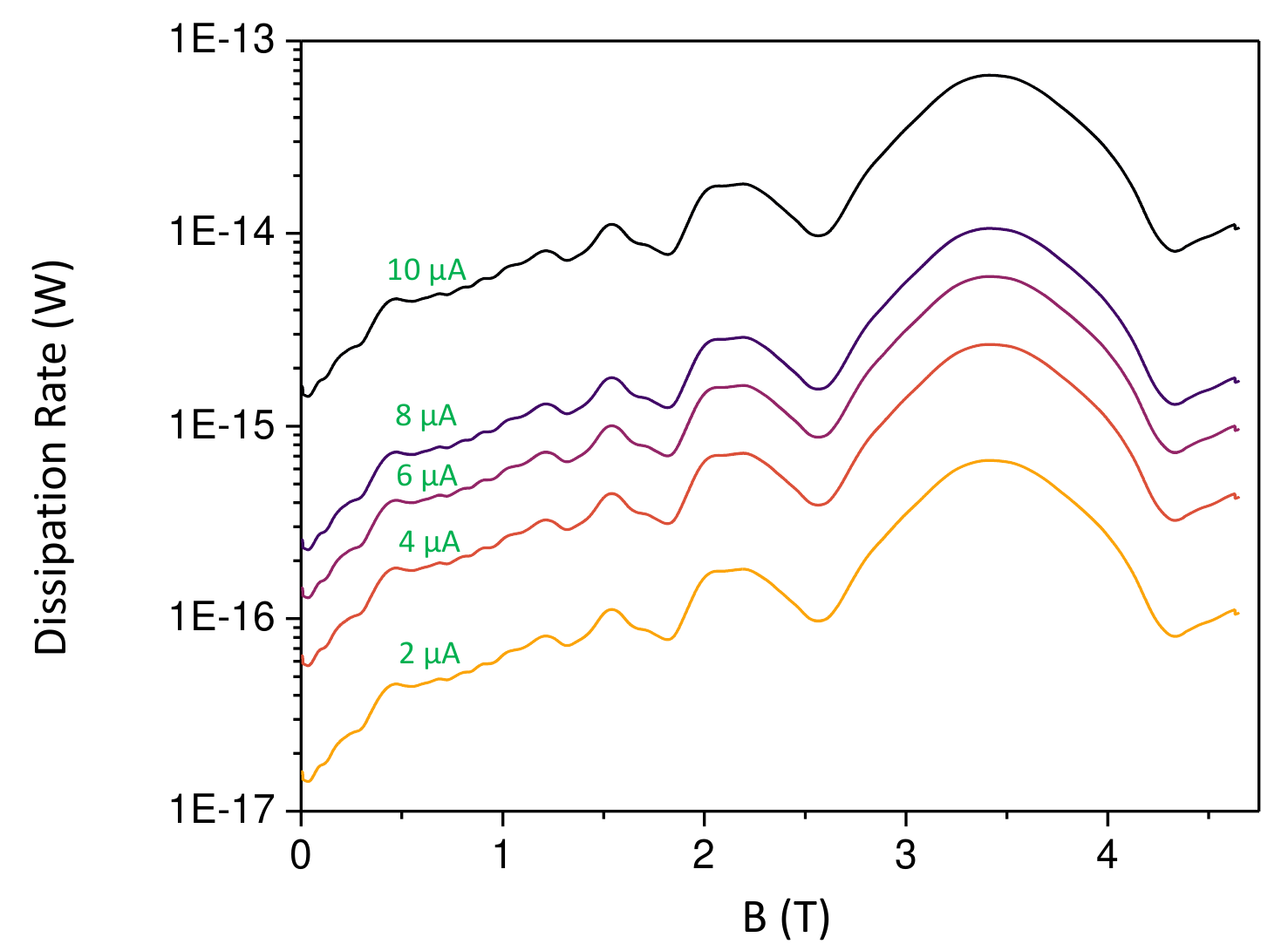}
              \captionsetup{labelformat=empty}
              \captionsetup{justification=raggedright,singlelinecheck=false}
              \caption{ Fig S1. \textbf{Dissipation rate as function of dc current:} Dissipation rate calculated using the equation $I_{dc} r_{xx}(0)/nA$, where $r_{xx}(0)$ is the  resistance of  the Shubnikov de Haas (SdH) oscillations at $I_{dc}=0$. Comparing the dissipation rate with previously reported values indicate the carrier temperature falls in range of $\sim 10$ K at a current of 10 $\mu$A and carrier density of 0.35 x 10$^{12}$ cm$^{-2}$[28, 29].}
      \end{figure*}

 \begin{figure*}
             \centering
              \includegraphics[scale=0.8]{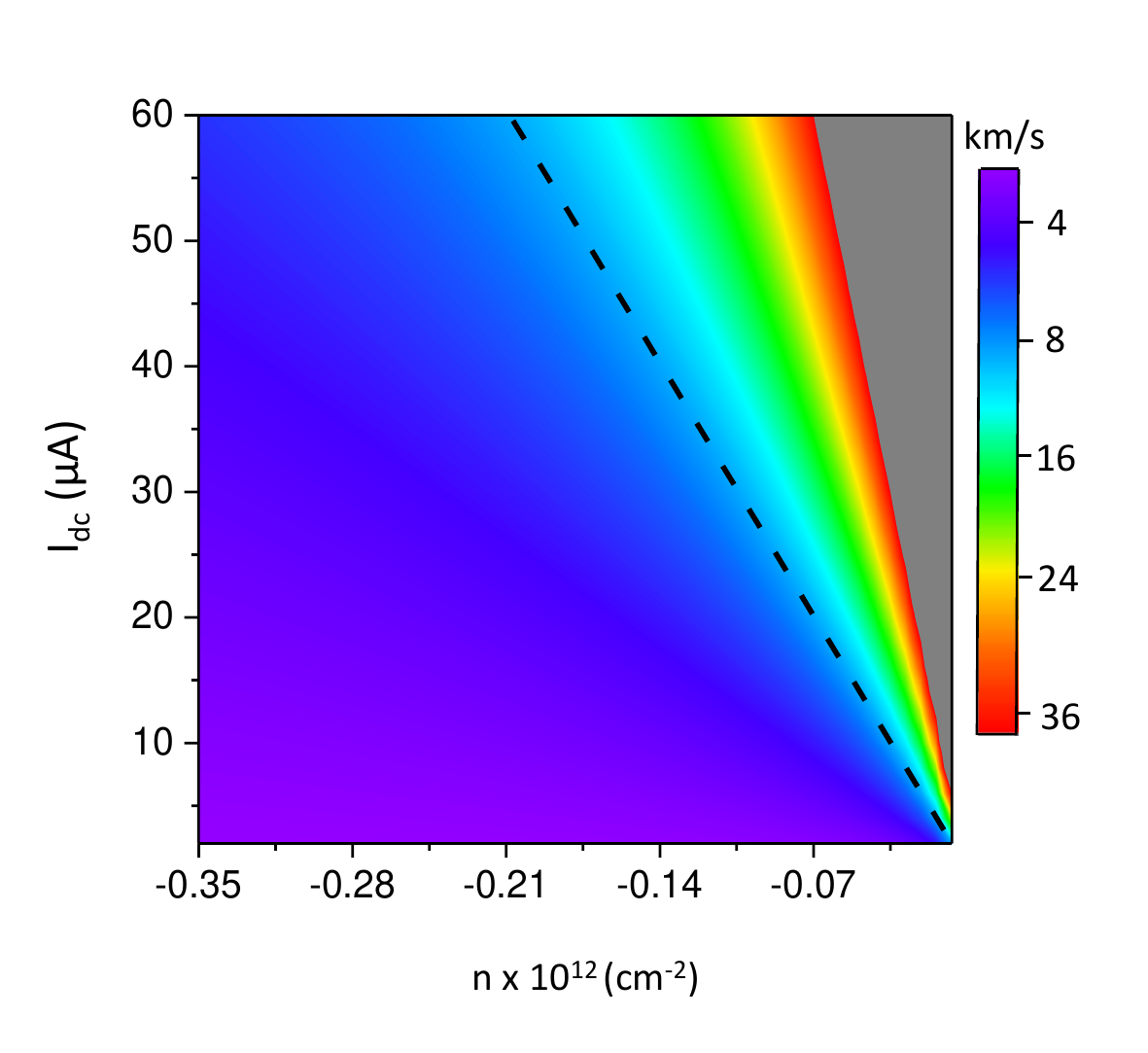}
              \captionsetup{labelformat=empty}
              \captionsetup{justification=raggedright,singlelinecheck=false}
              \caption{ Fig S2. \textbf{Drift velocity contour as function of dc current and carrier density:} Drift velocity induced due to the Hall field is calculated  as a function of dc current and carrier density in the hole doped region, close to the charge neutrality point. Within the experimental conditions, drift velocity $v_{d} = I_{dc}/new$ has been evaluated for $I_{dc}$ ranging between 0 to 60 $\mu$A  at 3 T, indicating that in the region where the new set of oscillations are seen, the drift velocity is of the same order of magnitude as the acoustic phonon velocities in graphene. Dashed line indicates a velocity of 10 km/s.}
      \end{figure*}

      \begin{figure*}
                \centering
                \includegraphics[scale=0.5]{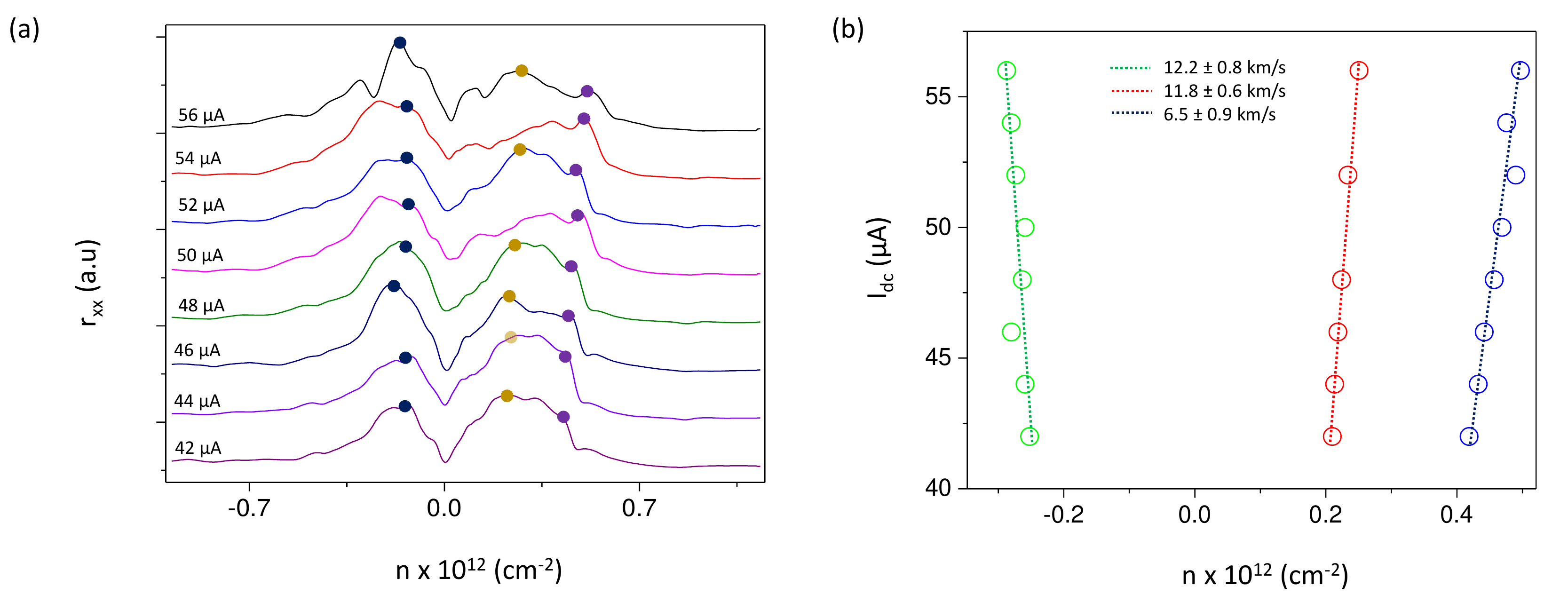}
                \captionsetup{labelformat=empty}
                 \captionsetup{justification=raggedright,singlelinecheck=false}
                \caption{Fig S3. \textbf{Effect of Magnetic field on novel magneto-oscillations near the charge neutrality point:} (a) Differential longitudinal magnetoresistance as function of carrier concentration and dc current at 5 T indicates the broadening of the oscillations near charge neutrality point which is consistent with the Landau-level broadening with increase in magnetic field. Consequently the two branches of phonon modes are not well resolved, in comparison to Fig.2 in the main text, although the general characteristics are retained. (b) Peak position as function of dc current and carrier density in the current ranging between 42 $\mu$A and 56 $\mu$A, as marked by the dots in (a). Velocities calculated from the dispersion of the peaks are 12.2 $\pm$ 0.8 km/s (hole side) and 11.8 $\pm$ 0.6 km/s and 6.5 $\pm$ 0.9 km/s (electron side) } 
        \end{figure*}

%\bibliography{5re}

\end{document}